\documentstyle[12pt]{article}
\begin{document}

\title{Extra $Z$ bosons and low-energy tests 
of unification}

\author{{\bf S.M. Barr and Almas Khan} \\
Bartol Research Institute \\ University of Delaware \\
Newark, Delaware 19716}

\date{\today}
             
\maketitle

\begin{abstract}
If there is an extra $U(1)$ gauge symmetry broken at low energies, then
it may be possible from the charges of the known quarks and leptons under 
this $U(1)$ to make inferences about how much gauge unification occurs 
at high scales and about the unification group. (For instance, there are 
certain observed properties of an extra $U(1)'$ that would be inconsistent 
with unification in four dimensions at high scales.) A general analysis
is presented.
Two criteria used in this analysis are (1) the degree to which
the generator of the extra $U(1)$ mixes with hypercharge, and (2) the ratio
of the extra $U(1)$ charge of the ``$10$" and the ``$\overline{5}$" of quarks
and leptons. 
\end{abstract}

\newpage

\section{Introduction}

It will never be possible to build an accelerator that reaches energies of
$10^{15}$ or $10^{16}$ GeV, so grand unified theories must be tested in
more indirect ways. One way is to look for very rare processes such as 
proton decay \cite{pdecay} or $n-\overline{n}$ oscillations \cite{nnbar}
or for cosmic relic superheavy particles such as magnetic monopoles
\cite{monopoles}. Another is to look 
for unification of parameters
by using the renormalization group to extrapolate to high scales, as has
been done for gauge couplings \cite{gaugeRGE} and may someday be possible 
for sparticle masses \cite{sparticleRGE}. 
A third way is find consequences of gauge symmetry
at low energy. An obvious example is the fact that the hypercharge
values of the Standard Model are those that would come from simple
$SU(5)$ or $SO(10)$ unification. Here we explore the possibility that the
charges of the known quarks and leptons under an ``extra" $U(1)$ group 
broken somewhat above the weak scale may enable us to infer something 
about gauge unification at high scales.

An obvious barrier to inferring anything about unification from 
gauge-charge values, is that the same assignments 
predicted by unification would in some instances also be required by
anomaly cancellation even without unification.  For instance, if one assumes 
that only the 
fermions of the Standard Model exist (without right-handed neutrinos)
and that hypercharge is family-independent, the anomaly conditions
fix the hypercharges of the quarks and leptons
uniquely to be the same values as would be predicted by $SU(5)$. 
(There are four anomaly conditions \cite{SManomaly}: 
$3^2 1_Y$, $2^2 1_Y$, $1_Y^3$, $1_Y$, in an obvious notation, the last being 
the mixed gravity-hypercharge anomaly.) Similarly, it might not always
be possible in the case of extra $U(1)$ charges
to distinguish the consequences of unification from those of anomaly
cancellation. (For discussions of anomaly constraints on extra $U(1)$
gauge groups see \cite{extraanomaly}.) 
That is one question we shall study in this paper.
We shall argue that one can distinguish in some circumstances, at least
in principle.

To see that there can be an ambiguity, consider the case of the extra
$U(1)$ contained in $SO(10)$. Let us
call this $U(1)_{X_{10}}$ and its generator $X_{10}$. On the fermions
of the Standard Model one has $X_{10} (e^+_L, Q_L, u^c_L, L_L, d^c_L, N^c_L) =
(1,1,1,-3,-3,5)$. On the other hand, with the same set of fermions 
and assuming that
charge assignments are family-independent, the six anomaly
conditions $3^2 1_X^2$, $2^2 1_X$, $1_Y^2 1_X$, $1_Y 1_X^2$, $1_X^3$, and
$1_X$ yield the same solution without unification. However, the anomaly 
conditions do not yield this solution uniquely, but only up to an arbitrary 
mixing with hypercharge. That is, the general solution of the anomaly 
conditions is $X = \alpha X_{10} + \beta (Y/2)$. In fact, it is easy to 
see that it is always the case that the six anomaly
conditions which have to satisfied by an extra $U(1)$ will allow the
generator of that $U(1)$ to have an arbitrary mixing with hypercharge. 
We shall exploit this fact: we shall see that under certain assumptions
the generator of an extra $U(1)$ {\it cannot} mix strongly with hypercharge
if there is gauge unification at high scales, whereas
it {\it can} and ``naturally" {\it ought} to mix strongly if there 
is no unification. The degree 
of mixing of the generator of the extra $U(1)$ with hypercharge will be 
one of the tools we shall use in our analysis.

We can also learn something about the degree of gauge unification at high
scales by comparing the extra-$U(1)$ charges
of the ``10" ($\equiv e^+_L, Q_L, u^c_L$) and the ``$\overline{5}$"
($\equiv L_L, d^c_L$). For example, in the simplest $SO(10)$ models 
one has $r \equiv X(10)/X(\overline{5}) =
-1/3$. Other schemes of unification give other characteristic values;
for example, if $SU(3) \times SU(2) \times U(1)_Y \times U(1)_X
\subset SU(6)$, then $r = -2$. On the other hand, partially unified or 
non-unified models can have values of this ratio that are not achievable
in any unified scheme. This will be the other tool of our analysis.

In our analysis, we do not use all the information about the extra $U(1)$
that may be obtained in principle from experiments. We are using only
the charges of the known quarks and leptons under the extra $U(1)$.
However, if the $Z'$ boson is 
actually produced in experiments, then almost certainly some of the 
extra fermions that must exist (to cancel the anomalies of the 
extra $U(1)$) will also be produced, since they are probably lighter than
the $Z'$. That would give additional information that would be helpful in 
making inferences about the degree of unification. This fact, of course, only
strengthens our main point, which is that from information about
extra $U(1)$ groups near the weak scale it is in principle possible 
to infer something definite about physics, and in particular about
unification, at very high scales. We also do not make use of conditions
on the charges of Higgs that arise from the requirement that the light
quarks and leptons be able to get realistic masses. Again, such considerations
might allow even stronger inferences to be made.

It should also be emphasized that we are making ``in principle" arguments 
in this paper.
We are not considering how to go about measuring the charges of the known
fermions under the extra $U(1)$, or concerned about the practical
feasibility of it. We are considering what can be measured ``in principle"
at low energies, and what can be inferred from it about very high energies.

We make use of assumptions of ``naturalness" in
several ways: (1) If no group-theoretical consideration or 
anomaly-cancellation condition forces it to do so, it is an unnatural 
fine-tuning for ratios of fermion charges under the extra $U(1)$ to 
be exactly equal to simple rational numbers like -2 or 1/2. (2) It is
difficult to make matter multiplets in unified theories have extreme mass
``splittings" (which is the basis of the well-known ``doublet-triplet splitting
problem" in unified theories). We assume that it is unnatural to have a
large number of such split multiplets besides the usual SM or MSSM
Higgs doublets. (3) It is assumed that if no symmetry or other principle makes
the mixing of the extra $U(1)$ generator with hypercharge small, it will not
be small.

The paper is organized as follows. In section 2, we shall explain our
assumptions, definitions, and notation and outline our results.
In sections 3 to 5 we shall explain the analyses that lead to those results.

\section{Assumptions, definitions, notation, and results}

We assume that the effective low energy theory below some scale $M_*$ 
has an $SU(3)_c \times SU(2)_L \times U(1)_Y \times U(1)'$ gauge symmetry,
which we will call $3211'$ for short. The $U(1)'$ is what we mean by the 
``extra" $U(1)$, and it is assumed broken at a scale $M'$ that is
above the weak scale, but close enough to it that it can
eventually be studied at accelerators. The generator of $U(1)'$ we call
$X'$ and the corresponding gauge boson $Z'$.

We will say that the theory 
is ``fully unified" if there is at some higher scale an effective 
four-dimensional theory with a simple gauge group $G$
such that $SU(3)_c \times SU(2)_L \times U(1)_Y \times U(1)'
\subset G$. We will say that it is ``partially unified"
there is an effective four-dimensional
theory with group $G \times H \supseteq G \times U(1)_X$, such that
$G$ contains $SU(3)_c \times SU(2)_L$ but does {\it not} contain
both low energy abelian groups $U(1)_Y \times U(1)'$.
Finally, we will say that it is ``non-unified" if there
is no four-dimensional unification of even the $SU(3)_c \times SU(2)_L$.

In both fully unified models and partially unified models we may write 
the low energy group between the scales $M_*$ and $M'$
as $SU(3)_c \times SU(2)_L \times U(1)_{Y_5} \times U(1)_X$, where
$SU(3)_c \times SU(2)_L \times U(1)_{Y_5} \subset SU(5) \subseteq G$, 
and $U(1)_X$ commutes with $SU(5)$. At $M'$ the breaking down to
the Standard Model group
can happen in two ways: (a) The generator $X$ is broken and $Y_5$
is left unbroken, in which case obviously $Y = Y_5$ and $X' = X$. This we call
``ordinary" or ``non-flipped" breaking. 
Or (b) both $X$ and $Y_5$ are broken at $M'$, 
leaving unbroken a hypercharge that is a linear
combination of $Y_5$ and $X$. Then we have $Y/2 = a Y_5/2
+ b X$ and $X'$ is the orthogonal linear combination
of $Y_5/2$ and $X$. This we call ``flipped breaking", as it
happens in ``flipped $SU(5)$" models (among others) \cite{flipped}. 

For convenience we will denote the set of multiplets $(e^+_L, Q_L, 
u^c_L)$ by ``10" and $( L_L, d^c_L )$ by ``$\overline{5}$" 
(with quotation marks) whether or not there is actually any $SU(5)$ 
unification. By the notation $\overline{X}$ we mean any
generator that has equal values for all the multiplets in ``10"
and equal values for the multiplets in ``$\overline{5}$". We will call
the ratio of these values $r$. That is, $r \equiv \overline{X}$(``10")/
$\overline{X}$(``$\overline{5}$"). 

Both unification (full or partial) and anomaly cancellation without
unification can lead to the result that $X'$ has the form
$X' = \alpha \overline{X} + \beta Y/2$. If $\beta/\alpha \neq 0$ and is
not small, we will say that $X'$ ``mixes strongly with hypercharge".
If $\beta/\alpha \ll 1$, we will say that there is small mixing. 
The degree of mixing with hypercharge is crucial to our analysis.

We will generally not assume anything about whether there is 
supersymmetry (SUSY). SUSY will not affect most of our analysis
if we make certain reasonable assumptions. SUSY would, of course, mean that
there would be Higgsinos that could be charged under the extra $U(1)$
and contribute to anomalies. However, these
contributions would typically cancel for the following reasons. Consider
the case of unification. The Higgs fields that get
vacuum expectation values (VEVs) at the weak scale, namely $H_u$ and $H_d$,
must then have color-triplet partners. These partners
must have masses much larger than $M'$ to avoid proton decay, and that
would require them to ``mate" with other triplets of opposite $X'$.
On the other hand, those Higgs fields that get VEVs of order $M'$ or 
larger must come paired with Higgs fields
that have opposite $X'$, generally, in order to avoid $D$-term breaking
of SUSY at large scales. 

\vspace{0.2cm}

\noindent
{\bf Why mixing with hypercharge is significant:}

\vspace{0.2cm}

In a model with no unification, there is no symmetry or other principle that
prevents $X'$, the generator of the extra $U(1)$, from mixing with hypercharge.
Anomaly-cancellation constraints never prevent this, and 
neither can the form of the Yukawa terms, since those terms must be invariant
under $U(1)_Y$ anyway. Therefore, one expects $X'$ to be of the form 
$X' = \alpha \overline{X} + \beta Y/2$, with $\beta/\alpha$ of order one. 

However, the situation is quite different in fully or partially unified 
models. As noted above, in such models one can write the low-energy 
group between the scales $M_*$ and $M'$ as $SU(3)_c \times SU(2)_L \times 
U(1)_{Y_5} \times U(1)_X$, where there is an $SU(5)$ that contains
$SU(3)_c \times SU(2)_L \times 
U(1)_{Y_5}$ and $U(1)_X$ commutes with that $SU(5)$. 
If there is ``ordinary" breaking of the extra $U(1)$ at $M'$, 
i.e. if only $U(1)_X$ 
breaks, then as noted before $X' = X$ and $Y = Y_5$, and so $U(1)'$
commutes with $SU(5) \supset SU(3)_c \times SU(2)_L \times
U(1)_Y$. That would mean that $X'$ does not mix with $Y$.

In other words, the unified group protects $X'$ from mixing with $Y$. However, 
because the unified group is broken at $M_{GUT}$, radiative effects 
induce a slight mixing below $M_{GUT}$.
In particular, a small effective mixing will in general be produced by
renormalization of the gauge kinetic 
terms \cite{kinetic}: one-loop diagrams produce a term
of the form $\epsilon F^{\mu \nu}_{(Y_5)} F_{(X) \mu \nu}$, which upon
bringing the gauge kinetic terms to canonical form shifts the gauge
fields to produces an effective
mixing of $X'$ and $Y$ with $\beta/\alpha \sim \epsilon$. If the 
particles going 
around the loop form complete, degenerate $SU(5)$ multiplets, then these 
diagrams vanish by the tracelessness of $Y_5$. However, split $SU(5)$
matter or Higgs
multiplets that contain both Weak-scale and superheavy components (like the
Higgs multiplets that break the Weak interactions) will give a contribution
to $\epsilon$ that is of
order $\frac{g_1 g_X}{16 \pi^2} \ln (M^2_{GUT}/M^2_W) \sim \frac{\alpha}{4 \pi}
\ln (M^2_{GUT}/M^2_W)$. For a Higgs doublet in a typical grand unified model,
like $SO(10)$, this will be about $0.02$. Thus, typically, in fully unified 
or partially unified models there is small mixing with hypercharge (of 
order a few percent).
Of course, in principle, large numbers of split multiplets
all contributing to $\epsilon$ with the same sign could exist and
produce strong mixing of $X'$ with hypercharge. However, it is notoriously
difficult to produce split multiplets naturally in unified models (hence
the ``doublet-triplet splitting problem"). The naturalness problem is 
compounded the more such split multiplets there are. It therefore seems 
quite unlikely that there would be large numbers of such multiplets. 
Even if there were,
one might expect that their light components would have mass near or 
below $M'$, where they could be observed and their effect on mixing
could be calculated and thus taken into account. Nevertheless, one 
cannot rule out the possibility that many split multiplets exist whose 
lighter components are at some inaccessible intermediate scale. 
However, this seems a highly artificial possibility.

The basic pattern, then, is simple: in non-unified models $X'$ is
expected to mix strongly with $Y$, whereas in partially unified or
fully unified models with ordinary breaking of the extra $U(1)$ the
mixing should be small (of order a few percent).

Matters are made slightly more complicated by the possibility in certain
cases of breaking at $M'$ that is not ``ordinary", as we shall now see. 
Suppose that
the groups $U(1)_{Y_5}$ and $U(1)_X$ both break at $M'$ leaving
unbroken $U(1)_Y$, with $Y/2 = a Y_5/2 + b X$, where $a,b \neq 0$. 
This implies that the broken generator $X'$ also is a linear combination
of $Y_5/2$ and $X$, and therefore of $Y/2$ and $X$, i.e. just
what we mean by ``mixing with hypercharge". Now consider what follows from
the requirement that the quark doublet $Q_L$ and the lepton doublet $L_L$ 
come out with the correct hypercharges. Since $Q_L$ has to be in the ${\bf 10}$
of $SU(5)$, it has $Y_5/2 = 1/6$. Call its $X$ value $x$.
Then one has $1/6 = a(1/6) +bx$. The lepton doublet must be in
the $\overline{{\bf 5}}$ of $SU(5)$ and thus have $Y_5/2 = -1/2$. Call
its $X$ value $x/r$. Then one has $-1/2 = a(-1/2) + bx/r$. Combining
these two equations gives $0 = b (3 + 1/r) x$. Since $b \neq 0$ 
there are only two possibilities. The first possibility is that 
$r = -1/3$, which corresponds to the $X$ charges of ${\bf 10}$ and
$\overline{{\bf 5}}$ being in the ratio $1$ to $-3$, as in $SO(10)$
models (but, as we shall see, not only in $SO(10)$ models). This leads 
to the well known ``flipped" breaking. This $r=-1/3$ case is
very special and has to be 
treated separately. We shall see that it still only produces ``small
mixing with hypercharge" in fully unified models, but can produce
``strong mixing with hypercharge" in 
partially unified models. 

The second possibility is
that $x =0$, i.e. the $X$ charge vanishes on both the
quark doublet and lepton doublet. In a fully unified or partially
unified model this means that $X$ vanishes on all the known
quarks and leptons. This also is a special case, which turns out
to be possible in partially unified models but not fully unified ones. 
It leads to mixing (i.e. $X' = \alpha X + \beta Y/2$), but since 
$X = 0$ on the known fermions, $X' = \beta Y/2$ on those fermions. 

We will
now simply state the results of our analysis and give that analysis
later.

\newpage

\noindent
{\bf Results of the analysis:}

\vspace{0.2cm}

We classify extra $U(1)$ models into seven types, based on whether
the generator $X'$ (corresponding to the massive $Z'$ boson)
has the form $X' = \alpha \overline{X} + \beta Y/2$, and the values of the 
parameters $\beta/\alpha$ and $r \equiv \overline{X}$(``10")/
$\overline{X}$(``$\overline{5}$"). The Classes are listed in an order
that moves generally from more unification to less.

\vspace{0.2cm}

\noindent
{\bf Class 1:} $X' = \alpha \overline{X} + \beta Y/2$, with $\beta/\alpha \ll 
1$ (``small mixing with hypercharge"), and
$r = -2$, 1/2, 4/3 or (perhaps) certain other simple rational values.

\vspace{0.2cm}

Such models are fully unified. If $r =-2$, then the full unification group 
is $SU(6)$ or some group containing it, either a larger unitary group or
$E_6$. If $r= 1/2$ or $4/3$, then the full unification group is either
$SO(10)$ or $E_6$. 
(In partially unified models, these values of 
$r$ would only result from tuning.)

\vspace{0.2cm}

\noindent
{\bf Class 2:} $X' = \alpha \overline{X} + \beta Y/2$, with $\beta/\alpha \ll 
1$ (``small mixing with hypercharge"),
and $r = +1$ or $-1/2$.

\vspace{0.2cm} 

Such models are either fully unified in $E_6$ or partially unified in
$G \times U(1)$, where $G \supseteq SO(10)$.

\vspace{0.2cm}

\noindent
{\bf Class 3:} $X' = \alpha \overline{X} + \beta Y/2$, with $\beta/\alpha \ll 
1$ (``small mixing with hypercharge"),
and $r$ is not equal to -2, 1/2, 4/3, 1, -1/2, -1/3 (i.e. the values 
characteristic of Classes 1, 2, and 4).

\vspace{0.2cm}

Such models are partially unified.

\vspace{0.2cm}

\noindent
{\bf Class 4:} $X' = \alpha \overline{X} + \beta Y/2$, with $\beta/\alpha \ll 
1$ (``small mixing with hypercharge"), and $r = -1/3$.

\vspace{0.2cm}

Such models are either fully unified (with gauge group $SO(10)$ or $E_6$)
or partially unified with ``ordinary" (i.e. non-flipped) breaking.
(The partial unification groups do not have to contain $SO(10)$ or $E_6$:
they may also be unitary groups.)

\vspace{0.2cm}

\noindent
{\bf Class 5:} $X' = \alpha \overline{X} + \beta Y/2$, $\beta \sim 1$
(``strong mixing with hypercharge"), and
$r = -1/3$. 

\vspace{0.2cm}

Such models are either partially unified with flipped 
breaking of $3211'$ to the Standard Model at $M'$,
or else they are non-unified, with specific extra fermions (i.e. fermions that
do not exist in the Standard Model).

\vspace{0.2cm} 

\noindent
{\bf Class 6:} $X' = \beta Y/2$.

\vspace{0.2cm}

Such models are either partially unified or non-unified, but cannot be
fully unified.

\vspace{0.2cm}

\noindent
{\bf Class 7:} $X' \neq \alpha \overline{X} + \beta Y/2$.

\vspace{0.2cm}

Such models are non-unified.

\section{Fully unified models}

As was shown in the previous section, in unified models there
will only be radiatively-induced (and typically small) mixing of $X'$ 
with hypercharge except in two special
cases: the case where $r =-1/3$ and the case where $X$ vanishes on
all the known quarks and leptons. If $r=-1/3$ in a fully unified model,
then the gauge group must be $SO(10)$ or $E_6$. This very special case 
will be treated at length in the next subsection. In the subsection after
that it will be shown that the case where $X$ vanishes on all the known
quarks and leptons (leading to $X' \propto Y/2$) cannot be realized in 
fully unified models. In the present subsection the general
case where $r \neq -1/3$ will be treated.

First, let us consider the simplest example of a fully unified
group: $SU(6)$. The simplest anomaly-free set of
$SU(6)$ fermion representations that gives one family consists of 
${\bf 15} + 2 \times (\overline{{\bf 6}})$.
(We shall also denote a $p$-index totally antisymmetric tensor
representation as $[p]$. So we could also write the anomaly-free set as
$[2] + 2 \times 
\overline{[1]}$.) The generator of $U(1)_X$ (in the fundamental
representation) is $X = diag(1,1,1,1,1,-5)$. Under the subgroup
$SU(5) \times U(1)_X$ the fermions of a family decompose into 
${\bf 10}^2 + {\bf 5}^{-4} + 2 \times (\overline{{\bf 5}}^{-1} + {\bf 1}^5)$. 
The effective theory 
below $M_*$ would have group $SU(3)_c \times SU(2)_L \times U(1)_{Y_5}
\times U(1)_X$. 

Since $X$ does not vanish on the known quarks and leptons, and 
$r \neq -1/3$, the analysis in the previous section tells us that there
is only the (typically small) radiatively-induced mixing of $X'$ 
with hypercharge, and therefore 
$X' = \alpha X + \beta Y/2$ with $\beta/\alpha \ll 1$, and $Y/2 = Y_5/2$. 
Thus, the Standard Model group is 
contained in the $SU(5)$; the ``10" $\equiv
(e^+_L, Q_L, u^c_L)$ is the ${\bf 10}$ of $SU(5)$; and the
``$\overline{5}$" $\equiv (L_L, d^c_L)$ is the
$\overline{{\bf 5}}$ of $SU(5)$. Consequently, the generator $X$ has 
equal values for all 
the multiplets in the ``10" and similarly for ``$\overline{5}$", and so we 
may put a bar over it and write $X' = \alpha \overline{X} + \beta Y/2$. 
Moreover, from the $X$ values of the
$SU(5)$ multiplets we see that $r \equiv \overline{X}$(``10")/$\overline{X}$
(``$\overline{5}$")$ = X({\bf 10})/X(\overline{{\bf 5}})
 = 2/(-1) = -2$. This model falls into Class 1.

It seems to be the case, as discussed in the Appendix, that $r=-2$ is the 
only value obtainable in realistic fully unified models based on the 
unitary groups, i.e. $SU(N)$. Thus $SU(N)$ full unification, as far as we 
can tell, leads always to models of Class 1. The value $r=-2$ can also 
arise in fully unified models based on $E_6$, since $E_6 \supset SU(6)$.
However, $r=-2$ does not seem to arise in fully unified models based
on $SO(10)$. Also, as we shall see 
in section 4, the value $r=-2$ does not seem to arise (except by artificial 
tuning of charge assignments) in partially unified models. 

There are some values of $r$, such as $1/2$ and $4/3$ that seem to arise 
only in fully unified models based on $SO(10)$. Consider $SO(10) \supset
SU(5) \times U(1)_X$, with each family containing a ${\bf 16} + {\bf 10}
= ({\bf 10}^1 + \overline{{\bf 5}}^{-3} + {\bf 1}^5) + 
( \overline{{\bf 5}}^2 + {\bf 5}^{-2})$. If the known quarks and leptons
are in the ${\bf 10}^1 + \overline{{\bf 5}}^2$, then $r=1/2$ results.
(It should be noted that in this case ``flipped" breaking is not possible,
and so there is not strong mixing with hypercharge as there can be in the
$r = -1/3$ case.) If each family consists of a ${\bf 16} + {\bf 45}$
of $SO(10)$, then the known quarks and leptons could be in a ${\bf 10}^{-4} +
\overline{{\bf 5}}^{-3}$ of $SU(5) \times U(1)_X$, yielding $r = 4/3$.
(This model has so many light multiplets that it can only narrowly 
escape a Landau pole at scales below the unification scale.)
Models with $r= 1/2$ and $r=4/3$ also fall into Class 1.

There are values of $r$, such as $1$ and $-1/2$ that can result 
from either full unification or partial unification.
These values arise from full unification in $E_6$ if $E_6 \supset SO(10) 
\times U(1)_X \supset
SU(3)_c \times SU(2)_L \times U(1)_Y \times U(1)_X$. Then, if a family is
a ${\bf 27}$, it decomposes under $SU(5) \times U(1)_X$ as
${\bf 10}^1 + \overline{{\bf 5}}^1 + {\bf 1}^1 + \overline{{\bf 5}}^{-2}
+ {\bf 5}^{-2} + {\bf 1}^4$ (where we use
$SU(5)$ multiplets as shorthand for the Standard Model multiplets). If
the known quarks and leptons are in ${\bf 10}^1 + \overline{{\bf 5}}^1$
then $r= +1$, and if they are in ${\bf 10}^1 + \overline{{\bf 5}}^{-2}$ 
then $r= -1/2$. However, these same values of $r$ can also
arise in partial unification based on $SO(10) \times U(1)_X \supset
SU(3)_c \times SU(2)_L \times U(1)_Y \times U(1)_X$, since anomaly 
cancellation alone is enough to fix the $U(1)_X$ charges to be the 
``$E_6$ values" if there are only ${\bf 16} + {\bf 10} + {\bf 1}$ in 
each family.
On the other hand, the values $r = +1$ and $-1/2$ do not 
arise in non-unified models (without artificial tuning of charge assignments). 
Models with $r=+1$ or $r=-1/2$ fall into Class 2.

The value $r = -1/3$, as noted before, is very special. It arises in
full unification based on $SO(10)$, but also, as we shall see in
later sections, it can arise naturally in both partially unified models and
non-unified models. Depending on how much mixing there is of the 
extra $U(1)$ charge $X'$ with hypercharge these models fall into 
Class 4 or 5. 

A general conclusion about fully unified models is that there is not strong 
mixing of the extra $U(1)$ charge with hypercharge (except in the 
somewhat artificial case that there are many highly split multiplets
that induce it raditively).

\vspace{0.2cm}

\noindent
{\bf An important special case: $r =-1/3$:}

In fully unified models, the case $r=-1/3$ arises only in $SO(10)$ or $E_6$.
Let us look at this special case more closely. (The present analysis will 
carry over almost completely also to the case $r=-1/3$ in partially
unified models.) 

Suppose one has 
$SU(3)_c \times SU(2)_L \times U(1)_{Y_5} \times U(1)_X \subset
SU(5) \times U(1)_X \subset SO(10)$. A family consists of
the $SU(5) \times U(1)_X$ representations ${\bf 10}^1 + 
\overline{{\bf 5}}^{-3} + {\bf 1}^5$. Let the covariant derivative 
contain the following combination of $U(1)$ gauge fields 

\begin{equation}
i D_{\mu} = i \partial_{\mu} + \left( g_1 \hat{\frac{Y_5}{2}} B_{1 \mu}
+ g_X \hat{X} B_{X \mu} \right) + ...,
\end{equation}

\noindent 
where the subscripts $1$ and $X$ refer
respectively to $U(1)_{Y_5}$ and $U(1)_X$, and we
denote by hats generators normalized
consistently in $SO(10)$, so that $tr_{16} \hat{\lambda}^2 = 2$. 
Then $\hat{\frac{Y_5}{2}} = \sqrt{\frac{3}{5}} \frac{Y_5}{2}$ and
$\hat{X} = \frac{1}{\sqrt{40}} X$ The flipped breaking
at $M'$ can be achieved by the VEV of an $SU(3)_c \times SU(2)_L$-singlet
field having $Y_5/2 = X = 1$ (such as exists in the spinor of $SO(10)$).
This leaves unbroken $Y/2 = \frac{1}{5}(- Y_5/2 + X)$. Therefore, 
in terms of the normalized generators, we may write

\begin{equation}
\begin{array}{l}
\hat{\frac{Y}{2}} = - \frac{1}{5} \hat{\frac{Y_5}{2}} + \frac{\sqrt{24}}{5}
\hat{X}.
\end{array}
\end{equation}

\noindent
The $U(1)$ charge that is orthogonal to this in $SO(10)$ is given by

\begin{equation}
\begin{array}{l}
\hat{\overline{X}} = \frac{\sqrt{24}}{5} \hat{\frac{Y_5}{2}} + \frac{1}{5}
\hat{X}.
\end{array}
\end{equation}

\noindent
Inspection of Eq. (1) shows that the massive gauge boson is

\begin{equation} 
Z'_{\mu} = \frac{\sqrt{24} g_1 B_{1 \mu} + 
g_X B_{X \mu}}{\sqrt{24 g_1^2 + g_X^2}},
\end{equation}

\noindent
and the gauge field $B_{\mu}$ of $U(1)_Y$ is the orthogonal combination

\begin{equation}
B_{\mu} = \frac{- g_X B_{1 \mu} + 
\sqrt{24} g_1 B_{X \mu}}{\sqrt{24 g_1^2 + g_X^2}},
\end{equation}

\noindent
Inverting Eqs. (4) and (5), Eq. (1) can be 
rewritten as

\begin{equation}
\begin{array}{l}
i D_{\mu} = i \partial_{\mu} + \frac{g_1 g_X}{\sqrt{24 g_1^2 + g_X^2}} 
\left[ - \hat{\frac{Y_5}{2}}
+ \sqrt{24} \hat{X} \right] B_{\mu} \\ \\ + 
\left[ \frac{\sqrt{24} g_1^2}{\sqrt{24 g_1^2 + g_X^2}} \hat{\frac{Y_5}{2}}
+ \frac{g_X^2}{\sqrt{24 g_1^2 + g_X^2}} \hat{X} \right] Z'_{\mu} + ... .
\end{array}
\end{equation}

\noindent
Then inverting Eqs. (2) and (3), this can be re-expressed as

\begin{equation}
\begin{array}{l}
i D_{\mu} = i \partial_{\mu} + 
\left( \frac{5 g_1 g_X}{\sqrt{24 g_1^2 + g_X^2}} \right)
\hat{\frac{Y}{2}} B_{\mu} \\ \\ + 
\left[ \frac{1}{5} \sqrt{24 g_1^2 + g_X^2} \hat{\overline{X}}
+ \frac{\sqrt{24}}{5} \frac{g_X^2 - g_1^2}{\sqrt{24 g_1^2 + g_X^2}} 
\hat{\frac{Y}{2}} \right] Z'_{\mu} ... .
\end{array}
\end{equation}

\noindent
Note that $B$ couples to hypercharge, as it should, and 
$Z'$ couples to a combination of $\overline{X}$ 
and hypercharge. Let us see what $\overline{X}$ is. 
It is convenient to normalize it as $\overline{X} =
\sqrt{40} \; \hat{\overline{X}} = \frac{1}{5}(24 \frac{Y_5}{2} + X)$. 
The charges of the known
quarks and leptons under $Y_5$, $X$, $Y/2$, and $\overline{X}$
are given in Table I. 

$$
\begin{array}{|c|c|cc|cc|c|}\hline
& & & & & & \\
{\rm field} & SU(5) & Y_5/2 & X & Y/2 & \overline{X} & \\
& & & & & & \\ \hline
N^c & {\bf 10} & 1 & 1 & 0 & 5 & {\rm ``1"} \\
Q & {\bf 10} & \frac{1}{6} & 1 & \frac{1}{6} & 1 & {\rm ``10"} \\
d^c & {\bf 10} & - \frac{2}{3} & 1 & \frac{1}{3} & -3 & 
{\rm ``\bar{5}"} \\
L & \overline{{\bf 5}} & - \frac{1}{2} & - 3 & - \frac{1}{2} & -3 & 
{\rm ``\bar{5}"} \\
u^c & \overline{{\bf 5}} & \frac{1}{3} & -3 & - \frac{2}{3} & 1 & 
{\rm ``10"} \\
e^+ & {\bf 1} & 0 & 5 & 1 & 1 & {\rm ``10"} \\ \hline
\end{array}
$$

\noindent
{\bf Table I:} The charges are related by $Y/2 = \frac{1}{5} [-(Y_5/2) + X]$
and $\overline{X} = \frac{1}{5} [24 (Y_5/2) + X]$.  

\vspace{0.5cm}

One sees that the ``10" $\equiv (e^+, Q, u^c)$ does not
coincide with the ${\bf 10}^1$ of $SU(5) \times U(1)_X$, and the
``$\overline{5}$" $\equiv (L, d^c)$ does not coincide with the 
$\overline{{\bf 5}}^{-3}$ of $SU(5) \times U(1)_X$ (though there is
another $SU(5) \times U(1)$ subgroup of $SO(10)$ of which they 
{\it are} multiplets). This is just the well-known phenomenon of flipping.
However, note that the generator $\overline{X}$ does have equal values 
for all the multiplets in the ``10" and equal values for all the 
multiplets in the ``$\overline{5}$". which is why we have denoted it
with a bar, consistent with the notation we explained in the previous
section. Thus, the generator $X'$ to which $Z'$ couples can 
be written

\begin{equation}
\begin{array}{ccl}
X' & = & \alpha \hat{\overline{X}} + \beta \hat{\frac{Y}{2}} \\ & & \\
& = & \frac{1}{\sqrt{40}} \alpha \overline{X} + \sqrt{\frac{3}{5}} 
\beta \frac{Y}{2},
\end{array}
\end{equation}

\noindent
where from Eq. (7) one has

\begin{equation}
\begin{array}{c}
\beta/\alpha = \frac{\sqrt{24}(g_X^2 - g_1^2)}{24 g_1^2 + g_X^2}.
\end{array}
\end{equation}

\noindent
In other words, there is ``mixing with hypercharge". If the couplings
$g_1$ and $g_X$ were equal at $M'$, the expression in Eq. (9) would 
vanish. Of course, these couplings are
equal at the scale where $SO(10)$ breaks; however, they will in general
run slightly differently between $M_{SO(10)}$ and $M'$ due primarily
to the Higgs
contributions to the beta functions. The known quarks and leptons 
do {\it not} make $g_1$ and $g_X$ run differently at one loop, because
they form complete $SO(10)$ multiplets. (Remember that the $N^c$
have masses of order $M'$ since they are protected by $U(1)_X$. It
is possible for some other, $SO(10)$-singlet fields to play the role
of superheavy right-handed neutrinos for the see-saw mechanism.) 
The Higgs-boson multiplets' contribution being relatively small, one expects
that the contribution to $\beta/\alpha$ from Eq.(9) will
be rather small. Indeed, in typical cases it turns out to be a few percent.
Of course, there is also the contribution to $\beta/\alpha$ coming from
the radiatively-induced gauge kinetic mixing discussed earlier, which is also a 
few percent typically. So in fully unified models, the mixing with hypercharge 
is small whether the breaking is flipped or ordinary. We shall see
in the next section that this is not the case in partially unified models.

\vspace{0.2cm}

\noindent
{\bf Another special case: $X' \propto Y/2$ and why it is impossible:}

As noted in section 2, mixing of $X'$ with hypercharge is possible
in unified models if $X$ vanishes on the known quarks and leptons.
Then $Y/2 = \alpha Y_5/2 + \beta X \propto Y_5/2$, which, of course, 
would give the right hypercharge assignments. This would be interesting 
as it would mean that $X'$ (which is also a linear combination of $Y_5/2$ 
and $X$) would 
have values for the known quarks and leptons proportional to their
hypercharges --- the defining characteristic of models in Class 6.
However, we will now show that in a {\it fully} unified model this 
possibility cannot be realized (though it can be realized
in partially unified models). The reason is that in a fully unified model
there are in general ``extra" non-singlet fermions in each family whose
existence is compelled by the fact that the multiplets of the full 
unification groups are large. It turns out that if $X$ vanishes on the known
fermions then the ``extra" fermions end up being chiral under 
hypercharge and electric charge and thus cannot obtain mass.
We will illustrate this with some examples, and it will be obvious that it 
generalizes.

\vspace{0.2cm}

\noindent
{\bf Example 1:}  Suppose that $SU(3)_c \times SU(2)_L \times U(1)_{Y_5}
\times U(1)_X \subset SU(5) \times U(1)_X \subset SU(7)$, with the
generator $X$ being $X =
diag(0,0,0,0,0,\frac{1}{2}, -\frac{1}{2})$.  An anomaly-free set
that gives one family is $[2] + 3 \times \overline{[1]} =
{\bf 21} + 3 \times (\overline{{\bf 7}})$. Under $SU(5) \times U(1)$ this
decomposes to ${\bf 10}^0 + {\bf 5}^{1/2} + {\bf 5}^{-1/2} + 
{\bf 1}^0 + 3 \times (\overline{{\bf 5}}^0 + {\bf 1}^{1/2} + {\bf 1}^{-1/2})$.
By assumption, $X$ vanishes on the known quarks and leptons, which therefore
consist of $({\bf 10}^0 + \overline{{\bf 5}}^0)$, and the remaining
fermions ${\bf 5}^{1/2} + {\bf 5}^{-1/2} + \overline{{\bf 5}}^0 +
\overline{{\bf 5}}^0$ etc. must ``mate" to obtain masses large enough that they
have not been observed. However, the hypercharge of the standard model
is, by assumption, a non-trivial
linear combination of $Y_5/2$ and $X$. Therefore it
is clear that the fields in ${\bf 5}^{1/2} + {\bf 5}^{-1/2}$ do
not have hypercharges that are opposite to the hypercharges of the fields 
in $\overline{{\bf 5}}^0 +
\overline{{\bf 5}}^0$, and consequently they do not have opposite electric
charges either. They are prevented from
acquiring mass unless electric charge breaks. Moreover, the
residual light fermions in ${\bf 5}^{1/2} + {\bf 5}^{-1/2}$ will have
exotic hypercharges.

\vspace{0.2cm}

\noindent
{\bf Example 2:} The previous example can easily be generalized to $SU(N)$.
Consider $SU(3)_c \times SU(2)_L \times U(1)_Y \times U(1)_X
\subset SU(5) \times U(1)_X \subset SU(N)$. Let $X$ (in the fundamental
representation) be given by 
$diag(0,0,0,0,0,\frac{1}{2}, -\frac{1}{2},0, ... ,0)$, where
the first five entries correspond to the $SU(5)$ that contains
$SU(3)_c \times SU(2)_L$.
Let the fermions be in totally antisymmetric tensor representations:
$n_1 \times [1] + n_2 \times [2] + n_3 \times [3] + ... $. 
An antisymmetric tensor representation
decomposes under the $SU(5) \times U(1)$ subgroup as follows.

\begin{equation}
\begin{array}{ccl}
[p] & \longrightarrow & 
\left[ \left( \begin{array}{c} N-7 \\ p-4 \end{array} \right) + 
\left( \begin{array}{c} N-7 \\ p-6 \end{array} \right) \right] \times 
\overline{{\bf 5}}^0 
+ \left( \begin{array}{c} N-7 \\ p-5 \end{array} \right) \times
\left( \overline{{\bf 5}}^{1/2} + \overline{{\bf 5}}^{-1/2} \right) \\
& + & 
\left[ \left( \begin{array}{c} N-7 \\ p-3 \end{array} \right) + 
\left( \begin{array}{c} N-7 \\ p-5 \end{array} \right) \right] \times 
\overline{{\bf 10}}^0 +
\left( \begin{array}{c} N-7 \\ p-4 \end{array} \right) \times 
\left( \overline{{\bf 10}}^{1/2} + \overline{{\bf 10}}^{-1/2} \right) \\
& + &
\left[ \left( \begin{array}{c} N-7 \\ p-2 \end{array} \right) + 
\left( \begin{array}{c} N-7 \\ p-4 \end{array} \right) \right] \times 
{\bf 10}^0 + 
\left( \begin{array}{c} N-7 \\ p-3 \end{array} \right) \times 
\left( {\bf 10}^{1/2} + {\bf 10}^{-1/2} \right) \\
& + & 
\left[ \left( \begin{array}{c} N-7 \\ p-1 \end{array} \right) + 
\left( \begin{array}{c} N-7 \\ p-3 \end{array} \right) \right] \times 
{\bf 5}^0 +
\left( \begin{array}{c} N-7 \\ p-2 \end{array} \right) \times
\left( {\bf 5}^{1/2} + {\bf 5}^{-1/2} \right) \\
& + & {\rm singlets}.
\end{array}
\end{equation}

\noindent
The known Standard Model families must consist, by assumption, of
$3 \times ({\bf 10}^0 + \overline{{\bf 5}}^0)$. The remaining fermions, 
if they are 
to get mass, must be vectorlike
under $U(1)_{Y_5} \times U(1)_X$. (Otherwise, their masses would break 
electric charge, as we have seen.) That means that
there must be equal numbers of $({\bf 10}^{1/2} + {\bf 10}^{-1/2})$ and of
$(\overline{{\bf 10}}^{1/2} + \overline{{\bf 10}}^{-1/2})$, and
similarly of $({\bf 5}^{1/2} + {\bf 5}^{-1/2})$ and of
$(\overline{{\bf 5}}^{1/2} + \overline{{\bf 5}}^{-1/2})$. These two conditions 
give, respectively,

\begin{equation}
\begin{array}{l}
\sum_{p} n_p \left( \begin{array}{c} N-7 \\ p-3 \end{array} \right) =
\sum_{p} n_p \left( \begin{array}{c} N-7 \\ p-4 \end{array} \right), \\
\sum_{p} n_p \left( \begin{array}{c} N-7 \\ p-2 \end{array} \right) =
\sum_{p} n_p \left( \begin{array}{c} N-7 \\ p-5 \end{array} \right).
\end{array}
\end{equation}

\noindent
However, these imply that the number of ${\bf 10}^0$ minus
the number of $\overline{{\bf 10}}^0$, i.e. the number of families,
must vanish:

\begin{equation}
n_{fam} = \sum_{p} n_p \left[ \left( \begin{array}{c} N-7 \\ p-2 \end{array} 
\right) + \left( \begin{array}{c} N-7 \\ p-4 \end{array} \right)
- \left( \begin{array}{c} N-7 \\ p-3 \end{array} \right)  
- \left( \begin{array}{c} N-7 \\ p-5 \end{array} \right) \right] = 0.
\end{equation}

We believe that this generalizes to all other types of representations,
other full-unification groups, and other $U(1)_X$ subgroups. 

\section{Partially unified models}

We have defined a partially unified model to be one where the group
$3211'$ describing physics below $M'$ is embedded as follows:
$SU(3)_c \times SU(2)_L \times U(1)_Y \times U(1)_X \subset SU(5)
\times U(1)_X \subseteq G \times U(1)_X \subseteq G \times H$, 
where $G$ is a simple group.
The same reasoning as for fully unified groups shows that $X$ does mix
strongly with hypercharge except in two special cases: (a) $r=-1/3$ and 
the breaking at $M'$ happens in a ``flipped" way, or
(b) $X$ vanishes on the known quarks and leptons. The reason, again,
is that except for these two special cases strong mixing of $X'$ with
hypercharge will cause the hypercharges of the known quarks and leptons to
come out wrong. The value $r=-1/3$ arises in the simplest
$SO(10)$ models, and so we will call models with $r = -1/3$ 
``$SO(10)$-like", even though, as we shall see, they may be based on
other groups (including unitary ones), both partially unified and non-unified.

\vspace{0.2cm}

\noindent
{\bf The ``$SO(10)$-like" and flipped special case:} Consider a model with 
group $SU(5) \times U(1)_X$ and fermion multiplets (per family) of 
${\bf 10}^a +
\overline{{\bf 5}}^b + {\bf 1}^c$. Then there are three anomalies that
must be satisfied by the $X$ charges: $5^2 1_X$, $1_X^3$ and $1_X$. These give 
the unique solution (up to overall normalization) $(a,b,c) =
(1,-3,5)$. (As always, we assume that $X$ is family-independent.)
These are the same charges that would be obtained if $SU(5) \times
U(1)_X$ were embedded in $SO(10)$. We will therefore call such models
``$SO(10)$-like". The analysis given in Eqs. (1)--(9) of what
happens if the $U(1)_X$ is broken in a flipped manner applies here
as well, except that the gauge coupling of $U(1)_{Y_5}$ is not unified with
that of $U(1)_X$. Consequently, what we called $g_1$ and $g_X$
are not related, and there is no reason for the parameter
$\beta/\alpha$ given in Eq. (9) to be small. Rather, one expects it
to be of order one, typically. This gives models of Class 5, then, rather
than Class 4. 

It is worth noting that one can get $SO(10)$-like models with other
choices of fermion content and other partial unification groups.
For example, in $SU(5) \times U(1)_X$, if there are (per family)
${\bf 10}^a + \overline{{\bf 5}}^b + {\bf 1}^c + {\bf 1}^d$, the unique
solution (up to overall normalization) is $(a,b,c,d) = (1,-3,5,0)$. 
Note that the ${\bf 1}^0$ could
play the role of right-handed neutrino with superheavy mass, giving
realistic see-saw masses for the light neutrinos. later we shall see an 
$SO(10)$-like model resulting from unitary groups like $SU(6) \times U(1)$.

\vspace{0.2cm}

\noindent {\bf The $X' \propto Y/2$ special case:}
This special case can be realized in partial unification without
producing massless fermions with exotic charges --- in fact quite, 
trivially. For example,
let the only quarks and leptons be in $3 \times ({\bf 10}^0 + 
\overline{{\bf 5}}^0 + {\bf 1}^0)$ of $SU(5) \times U(1)_X$, and let 
some Higgs field (for example ${\bf 10}_H^q$) break both
$Y_5$ and $X$ at $M'$, leaving unbroken $SU(3) \times SU(2) \times
U(1)$. There is no problem here with extra quark and lepton multiplets that
have chiral values of hypercharge and electric charge which prevent
them from obtaining mass, since unlike the
fully unified case there is here
no larger simple group containing $SU(5) \times U(1)$ that 
implies their existence. Thus models of Class 6 can arise from partial 
unification.

\vspace{0.2cm}

\noindent
{\bf The general case of no mixing:}
Turning now to the more generic cases where $X'$ does {\it not}
mix strongly with hypercharge,
we will show that the partially unified models can be distinguished from 
the fully unified ones by the fact that they generally give different values of
$r$. It is simplest to consider a few examples. 

Consider, first, a model with group $SU(5) \times U(1)_X$ and fermions
content (per family) consisting of ${\bf 10}^a + \overline{{\bf 5}}^b
+ \overline{{\bf 5}}^c + {\bf 5}^d + {\bf 1}^e$. There is a unique
solution of the anomaly conditions
(up to interchange of the two $\overline{{\bf 5}}$'s and overall 
normalization): $(a,b,c,d,e) = (1,-3,x,-x,5)$, with $x$ undetermined.
If one takes a family
to consist of ${\bf 10}^1 + \overline{{\bf 5}}^{-3} + {\bf 1}^5$, one has
an $SO(10)$-like model. However, it is possible that a family consists
of ${\bf 10}^1 + \overline{{\bf 5}}^{x} + {\bf 1}^5$. In this case, one
has $r = \overline{X}({\bf 10})/\overline{X}(\overline{{\bf 5}})
= X({\bf 10})/X(\overline{{\bf 5}}) = 1/x$. This can be any number;
anomaly cancellation leaves it completely undetermined. It therefore has
no reason to be equal to one of the characteristic values (like -2 and 
-1/3) that occur in fully unified models. Therefore, such a model would
be in Class 3. 

If one requires in this latter model (where a family is in 
${\bf 10}^1 + \overline{{\bf 5}}^{x} + {\bf 1}^5$) that the light Higgs
doublets $H_d$ and $H_u$ have opposite $X$, and further that the
quark and lepton masses all come from dimension-four Yukawa terms,
then it would force $X(H_u) = -2$, $X(H_d) = +2$, and $x = -3$, 
giving an $SO(10)$-like model. 
However, it is also possible that $H_u, H_d$ have $X = -2, +2$, but
that the Yukawa terms have the form: $({\bf 10}^1 {\bf 10}^1) {\bf 5}^{-2}_H$
and $({\bf 10}^1 \overline{{\bf 5}}^x) \overline{{\bf 5}}_H^2 
{\bf 1}^{-3-x}_H/M'$.
Note that there must, in any event, be a Higgs field ${\bf 1}_H^{3 +x}$ with
VEV of order $M'$ if the extra fermions 
$\overline{{\bf 5}}^{-3} + {\bf 5}^{-x}$
are to get large mass together. In fact, by integrating out these extra
fermions, it is 
possible to get just the dimension-five Yukawa term that we have written.
Thus, the Yukawa terms do not impose any {\it a priori} constraint on $x$ or
$r$. (This integrating out of the extra fermions to produce effective
dimension-five Yukawa terms for the known quarks and leptons will
produce some mixing between the $\overline{{\bf 5}}^{-3}$ and the 
$\overline{{\bf 5}}^x$. This mixing can be small, in which case
$r$ would be given very nearly by $1/x$ for all families. If, however,
this mixing is large, one would get a complicated pattern of couplings
to $Z'$ that would not in general be family-independent.)

The next example is one where anomalies actually force a unique
solution for $r$, but it does not come out to be one
of the characteristic values that arise from full unification.
Consider a model with gauge group $SU(5) \times U(1)$ and fermion
content (per family) of ${\bf 10}^a + \overline{{\bf 5}}^b  
+ {\bf 15}^c + \overline{{\bf 15}}^d$. This gives the solutions
$(a,b,c,d) = (4, -5, -1, 0)$ or $(4, -5, 0, -1)$. In either case,
$r = -4/5$. This model is also in Class 3.

These examples show that one can get values of $r$ in partially 
unified models that do not arise in fully unified models. 
Turning the question around, one can ask
whether partially unified models can give all the values of $r$
that do arise in fully unified models. Of course, one can get them in some
models by simply choosing the charges arbitrarily to have the right values,
which is a kind of fine-tuning. The question is whether anomaly 
cancellation without full unification can {\it force}
$r$ to have one of those values characteristic of full unification.
We have already seen that the answer is yes for the special cases 
$r=+1$ and $r=-1/2$ (which can come from $SO(10) \times U(1)_X$ as well
as $E_6$) and $r=-1/3$ (which can arise in many ways).
However, these $E_6$-like and $SO(10)$-like values are special in this regard. 
The value $r= -2$ characteristic of full-unification based on $SU(N)$
does not seem ever to be forced by anomaly cancellation in partially unified 
models.

We will now see why this the case by looking at a simple example.
We saw that the value $r=-2$ can arise in fully unified models based on
$SU(6)$. Can the value $r=-2$ be forced by
anomaly cancellation (plus the assumption of family-independence)
in a partially unified model? In the simplest $SU(6)$ model, a 
family consists of ${\bf 21} + 2 \times \overline{{\bf 7}}$, 
which decomposes under the $SU(5) \times U(1)_X$ subgroup into
${\bf 10}^2 + {\bf 5}^{-4} + 2 \times (\overline{{\bf 5}}^{-1} +
{\bf 1}^5)$, as noted before. One might take this set of $SU(5)$ 
representations and ask whether anomaly cancellation alone would force 
the same solution for the $U(1)_X$ charges. The answer is no. 
For the set ${\bf 10}^a + {\bf 5}^b +
\overline{{\bf 5}}^c + \overline{{\bf 5}}^d + {\bf 1}^e + {\bf 1}^f$,
the most general solution to the anomaly-cancellation conditions has two
parameters (not counting overall normalization). A simple one-parameter
subset of this solution is $(a,b,c,d,e,f) = (2,-4,-1+x, -1-x, 5+x, 5-x)$,
with $x$ arbitrary; so that $r= 2/(-1 \pm x)$ and can be anything. (Note that 
$x= \mp 5$ would give $SO(10)$-like charges, and $x=0$ gives $SU(6)$-like
charges.) Anomaly cancellation does not force a particular value of $x$.
The reason for this ambiguity lies in $E_6$. $E_6$ has the
chain of subgroups $E_6 \supset SU(6) \times SU(2) \supset SU(5) \times U(1)_6
\times SU(2) \supset SU(5) \times U(1)_6 \times U(1)_2$. The ${\bf 27}$
of $E_6$ decomposes into these $5 1_6 1_2$ multiplets:
${\bf 10}^{(2,0)} + {\bf 5}^{(-4,0)} + 
(\overline{{\bf 5}}^{(-1,+1)} + \overline{{\bf 1}}^{(-1,-1)}) +
({\bf 1}^{(5,+1)} + {\bf 1}^{(5,-1)})$. Clearly, any $U(1)$ generator that
is a linear combination of the generators of $U(1)_6$ and $U(1)_2$
will satisfy the anomaly cancellation conditions since $E_6$ is an 
anomaly-free group.
The undetermined parameter $x$ that we found in the $SU(5) \times
U(1)$ solution just reflects this ambiguity.

If one reduces the number of multiplets per family, one reduces
the number of unknowns and may get a unique solution to the anomaly
cancellation conditions; however, the unique solution 
will not be $r=-2$. In fact, if we remove 
one of the singlets one has the
first example in this subsection, for which anomaly cancellation gives  
${\bf 10}^1 + \overline{{\bf 5}}^{-x}
+ \overline{{\bf 5}}^{-3} + {\bf 5}^x + {\bf 1}^5$, so that 
either $r=-1/3$ or $r = 1/x$ (i.e. undetermined). If we remove a pair
of fundamental plus antifundamental, we have already seen that one gets
uniquely the $SO(10)$-like solution ${\bf 10}^1 + \overline{{\bf 5}}^{-3} + 
{\bf 1}^5 + {\bf 1}^0$. 

On the other hand, if one adds more $SU(5)$ multiplets per family,
it just increases the number of undetermined parameters, so that again
$r$ is not forced to be $-2$. Nor does going to larger partial unification
groups allow situations where anomalies force $r=-2$. Consider, for
instance, $SU(6) \times U(1)_1$ with fermions ${\bf 15}^a + 
\overline{{\bf 6}}^b +
\overline{{\bf 6}}^c + {\bf 1}^d$. The three anomaly conditions ($6^2 1$,
$1^3$, $1$) force the values $(a,b,c,d) = (1,1, -5, 9)$ (up to an overall
normalization). Under the subgroup $SU(5) \times U(1)_6 \times U(1)_1$ the
fermions of a family decompose into ${\bf 10}^{(2,1)} + {\bf 5}^{(-4,1)}
+ \overline{{\bf 5}}^{(-1,1)} + {\bf 1}^{(5,1)} + \overline{{\bf 5}}^{(-1,-5)}
+ {\bf 1}^{(5,-5)} + {\bf 1}^{(0,9)}$. The extra $U(1)_X$ is some linear
combination of $U(1)_6$ and $U(1)_1$. Which linear combination it is depends
on how the groups break at the scale $M_*$, and that in turn depends
on what kinds of Standard Model-singlet Higgs fields exist in the model.

Generally, the $U(1)_1$ charges of the Higgs fields are not constrained
by anomaly cancellation. (Even in supersymmetric models where the Higgs fields
have fermionic partners, these generally come in conjugate
pairs, for reasons explained above, and so their anomalies cancel.)
There is at least one Standard Model-singlet Higgs fields that must
appear in such a model, namely the one required to give mass to the ``extra"
$\overline{{\bf 5}} + {\bf 5}$ of quarks and leptons. There are two
possibilities: the ${\bf 5}^{(-4,1)}$ can either get mass with the 
$\overline{{\bf 5}}^{(-1,1)}$ or with the $\overline{{\bf 5}}^{(-1,5)}$.
In the former case, the required singlet Higgs is ${\bf 1}^{(5,-2)}_H$, 
in the latter it is ${\bf 1}^{(5,4)}_H$. In neither case does the singlet 
Higgs have to be the one responsible for breaking down to $3211'$ at $M_*$.
(These singlets could get VEVs much smaller than the scale $M_*$, with some 
other singlet doing the breaking at $M_*$.)
However, if either of them is the one responsible for the breaking to
$3211'$ at $M_*$, one easily sees that an $SO(10)$-like extra $U(1)$
is left unbroken. For example, if $\langle {\bf 1}_H^{(5,-2)} \rangle
\sim M_*$, then the unbroken generator is $X = (2 X_6 + 5 X_1)/9$, where
we have used a convenient normalization. This leads to the known quarks
and leptons having $X({\bf 10}^{(2,1)}) = 1$ and 
$X(\overline{{\bf 5}}^{(-1,-5)}) = -3$, giving $r = -1/3$. The other case
is similar. The reason that an $SO(10)$-like model results is simple:
the theory below $M_*$ has the fermion content per family
${\bf 10} + \overline{{\bf 5}} + {\bf 1} + {\bf 1}$, which we have
already seen to be forced by anomaly cancellation to give a $U(1)_X$ that is
$SO(10)$-like. 

One sees that partial unification can lead to $SO(10)$-like models with
$r = -1/3$, in which case the models are Class 4 or Class 5, depending 
on whether the breaking of $3211'$ at $M'$ happens in the ordinary or 
flipped manner.
It can also lead to $E_6$-like models $r = +1$ or $-1/2$, which fall into 
Class 2. Finally, it can lead to models with arbitrary values of $r$ not 
equal to any of the special values characteristic of full unification;
such models fall into Class 3.

\section{Non-unified models}

In non-unified models, there is no unification of the groups
$SU(3)_c \times SU(2)_L \times U(1)_Y \times U(1)'$ above the scale 
$M_*$, and therefore no group-theory constraints on the charge assignments of 
{\it either} $U(1)$ group. If the only constraints on these charges come
from anomaly cancellation, there is no guarantee in general that
the hypercharge assignments will come out correct or even that a $U(1)$ 
group will be left unbroken when the extra fermions acquire mass at
$M'$. However, if after the breaking at $M'$ only the Standard Model
quark and lepton multiplets remain light and a $U(1)$ is indeed left
unbroken, then anomaly cancellation guarantees that the charge assignments
of the light quarks and leptons under the unbroken $U(1)$ will 
correspond to the known hypercharges \cite{adh}.

\vspace{0.2cm}

\noindent
{\bf Obtaining the hypercharge group}

Let us illustrate some of these points with a simple example.
Consider first a model in which there are two extra singlets per family,
so that each family consists of $(Q, u^c, d^c, L, e^c, N, N')$. Let the 
gauge group be $SU(3)_c \times SU(2)_L \times U(1)_1 \times U(1)_2$, where
we label the abelian factors as we do because we do not yet know whether
hypercharge will emerge from the anomaly conditions. (Of course, we assume as
always that the gauge groups couple in a family-independent way.)

There are ten anomaly
conditions that constrain the abelian charge assignments:
$3^2 1_1$, $2^2 1_1$, $1_1^3$, $1_1$,
$3^2 1_2$, $2^2 1_2$, $1_2^3$, $1_2$, $1_1^2 1_2$, and $1_1 1_2^2$.
The first four anomaly conditions, which constrain only the $U(1)_1$ charge
assignments, force them to be of the form $X_1 = (1, -4+x_1, 2-x_1, -3, 
6-e_1-f_1, e_1, f_1)$, where we list them in the same order as we listed 
the multiplets above.
The cubic anomaly condition gives the relation $0= 6x_1(x_1-6)
+ (e_1+f_1)(e_1-6)(f_1-6)$. We have chosen to normalize these charges so that
$X_1 (Q) = 1$. The next four anomaly conditions, which constrain only
the $U(1)_2$ charge assignments, force them to be
$X_2 = (1, -4+x_2, 2-x_2, -3, 6-e_2-f_2, e_2, f_2)$, where
$0= 6x_2(x_2-6) + (e_2+f_2)(e_2-6)(f_2-6)$. Again, we have normalized these to 
make $X_2 (Q) = 1$. 
Finally, the remaining two anomaly
conditions ($1_1^2 1_2$ and $1_1 1_2^2$) give a pair of cubic equations
that must be satisfied by the parameters $x_1$, $e_1$, $f_1$, $x_2$,
$e_2$, and $f_2$. Altogether, then, there are six parameters that must
satisfy four non-linear equations. That means that there are two-parameter
families of solutions. We may take those parameters to be $e_1$ and $f_1$,
which are the charges under $U(1)_1$ of the ``extra" singlets $N$ and $N'$,
and these may take values in a finite range. 

At first glance, it is not obvious that in the general case
any linear comination of
$X_1$ and $X_2$ will give the Standard Model hypercharges for the known
quarks and leptons. However, it is not difficult to see that one linear
combination does and that it is easy to break $U(1)_1 \times U(1)_2$
down to it. For suppose that there is a singlet Higgs field $S$ that
has Dirac couplings to the extra singlet fermions: $h_{ij} (N_i N'_j) S$, where
$i,j$ are family indices. If $S$
obtains a VEV of order $M'$ it leaves one linear combination of
$X_1$ and $X_2$ unbroken. Since it also leaves only the quarks and leptons
of the Standard Model light, we know from anomaly cancellation that
the unbroken $U(1)$ must act on the known quarks and leptons as
the Standard Model hypercharge (up to
an overall nomalization, of course, that can be absorbed into
the gauge coupling.) 

On the other hand, suppose that the extra fermions got mass at the scale $M'$
from Majorana terms like $(N N) \langle S \rangle + (N' N') \langle S' \rangle$.
Then, unless the charge assignments were tuned to special values, 
no $U(1)$ group would be left unbroken below $M'$, so that the Standard Model
would not be reproduced.

\newpage

\noindent
{\bf Models that reproduce the Standard Model}

If one is dealing with a model that reproduces the Standard Model,
then we can write $U(1)_1 \times U(1)_2 = U(1)_{Y'} \times U(1)'$,
where $Y'$ equals the standard hypercharges on the known quarks and leptons.
(However, $Y'$ need not have the ``standard" values on extra fermions:
on them it can have any values consistent with their mass terms. The 
extra fermions, whose masses are of order $M'$ are obviously vectorlike
under $Y'$.) Thus, 
$Y'$ satisfies several anomaly conditions automatically (namely
$3^2 1_{Y'}$, $2^2 1_{Y'}$, $1_{Y'}^3$ and $1_{Y'}$), and we need only consider
six anomaly cancellation conditions for the extra $U(1)$:
$3^2 1_{X'}$, $2^2 1_{X'}$, $1_{Y'}^2 1_{X'}$, $1_{Y'} 1_{X'}^2$, 
$1_{X'}^3$, $1_{X'}$. 

If the only fermions are those of the Standard Model, and their
charges are assumed to be family-independent, then there are only four
unknowns, namely the ratios of the $X'$ charges of $Q$, $u^c$, $d^c$, $L$,
$e^+$. The only solution is hypercharge itself, i.e. $X' = \beta Y/2$.
Of course, the Higgs fields or other scalars that might exist
in the low energy theory can have arbitrary $X'$. Such models would fall
into Class 6. 

If there are additional fermion multiplets per family, then several 
possibilities exist, depending on what those fermions are. In some 
cases, the solutions still give $X' = \beta Y/2$ on the known fermions
and fall into Class 6.
In other cases, the solutions are $SO(10)$-like, in that case
$X' = \alpha X_{10} + \beta Y/2$, on the known fermions, where 
$X_{10} (e^+, Q, u^c, L, d^c) = (1,1,1,-3,-3)$. Such models fall
into Class 5, since the parameter $\beta$ has no reason to be small.
(The mixing with hypercharge in this case does not have to arise from
a simultaneous breaking of $Y_5$ and $X$ at $M'$, as in partially unified
or fully unified models, where $X$ is a generator that commutes with
an $SU(5)$ and is unmixed with hypercharge to begin with. In
non-unified models, the anomaly conditions permit arbitrary mixing with
hypercharge, and there is {\it nothing} to make that mixing small --- not even
assumptions about which Yukawa couplings and Higgs multiplets exist, since
all Yukawa couplings must conserve hypercharge.) Finally, there are cases,
where the solutions are messy and not of the form $X' = \alpha \overline{X} 
+ \beta$. These fall into Class 7.

Let us look at some simple examples. (A) If there exists in addition to
the known light fermions just one Standard
Model singlet per family (call it $N$), then the most general
solution is $SO(10)$-like: $X' = \alpha X_{10} + \beta Y/2$, and falls into 
Class 5.
(In this case there are five unknowns satisfying six equations, since the
overall normalization of the charges does not matter. The fact that a
non-trivial solution exists is explained, of course,  by the fact that 
the fermions in this case are able to fit into the spinor of $SO(10)$,
even though no $SO(10)$ actually exists in the model.)

(B) If the extra fermions per family are just 
two Standard Model singlets, $N$ and $N'$,
then there are {\it two} solutions to the anomaly conditions. 
One solution has $X' = \alpha X_N + \beta Y/2$, 
where $X_N$ is $+1$ and $-1$ on the two singlets
and vanishes on all other fermions. On the known fermions this gives
$X' = \beta Y/2$, and therefore falls into Class 6. The other solution has
$X' = \alpha X_{10} + \beta Y/2$, where $X_{10}$ has the values given above
for the known fermions and is $+5$ and $0$ on the two singlets $N$, $N'$. 
This is $SO(10)$-like and falls into Class 5.

(C) If the extra fermions per family are just three singlets, 
$N$,, $N'$ and $N^{\prime \prime \prime}$,
then the general solution is $X' = \alpha X_{10} + \beta Y/2 
+ \gamma X_N$, where $X_{10}$ has the values given above for the known
fermions and is $+5$, $0$, and $0$ on the three singlets $N$, $N'$,
$N^{\prime \prime \prime}$. The
generator $X_N$ is $+1$ and $-1$ on the two singlets that have
$X_{10} =0$ and vanishes on the other. This solution will fall into
Class 5. (We assume that coefficients such as $\alpha$ are not tuned to zero
if group theory or anomaly cancellation do not require it.)
If there are more than three singlets per family one gets solutions similar to 
those above.

(E) If the extra fermions per family are just a 
conjugate pair $R + \overline{R}$ of
non-singlet irreducible representations, then the solution is 
$X' = \alpha X_R + \beta Y/2$, where $X_R$ is $+1$ and $-1$ on the
conjugate pair and vanishes on the known fermions. This falls into Class 6.
(If $R$ has the same 
Standard Model charges as a known fermion, then there is
a solution trivially obtained from the previous by interchanging the two
multiplets. This would fall into Class 7.) 

(F) If the extra fermions per family are a conjugate pair and a singlet, 
$R + \overline{R} + N$, then there are two distinct solutions. One
is $SO(10)$-like:
$X' = \alpha X_{10} + \beta Y/2 + \gamma X_R$, in a notation that is
obvious. This falls into Class 5. The other solution is a messy 
two-parameter solution that falls into Class 7.

These simple examples show what kinds of possibilities exist. Of course,
generally speaking, the more extra fermions that exist, the more undetermined 
parameters will exist in the solution. Complicated cases 
will usually fall into Class 7.

\section{Conclusions}

We have argued that a discovery of an extra $Z$ boson can provide information 
that allows inferences about the degree of gauge unification at high scales.
For example, if there is strong mixing of
the generator of the $U(1)'$ with hypercharge and $r \neq -1/3$
(and if $X'$ is not proportional to $Y$), it would strongly disfavor
conventional four-dimensional unification of the Standard Model in a simple 
group. (However, it would not rigorously disprove it, given the
possibility, which seems artificial and hard to obtain naturally, that
there might be numerous highly split multiplets that induce $O(1)$
radiative mixing of $X'$ with $Y$.). 
As another example, if
$X' \propto Y$ it disproves ``full unification", i.e. the
unification of the Standard Model group {\it and} the $U(1)'$ in 
a single simple group. 

On the other hand, the discovery of certain patterns of
$U(1)'$ charge assignments would be strong evidence in favor of certain
specific kinds of gauge unification. For instance, finding $r=-2$ and
small mixing of the $U(1)'$ charge with hypercharge would strongly favor full
unification in a group that contained $SU(6)$, i.e. either a unitary
group of $E_6$. However, as far as such positive inferences go, there
is always the possibility that certain charges can take special values
purely by accident --- by fine-tuning, as it were. 

There remains much more to be done. First, we have not yet
succeeded in rigorously proving some of our conclusions even though
we have strong evidence for them, based on both partial proofs and
the working out of examples. Second, there is the question of
how much stronger the conclusions would be if one also used information
about the spectrum of extra light quarks and leptons and their charges
under both the Standard Model group and the $U(1)'$. Third, it may be
that extra $U(1)'$ charges in the original gauge basis may be
family-independent, but that family-dependence arises as a result
of symmetry breaking and mixing in the fermion mass matrices. We have
not addressed this case, but only cases where the observed charge
assignments are family-independent. 

Of course, it is likely that there is no extra gauge symmetry at
low energies. However, if there is, it would prove to be a potent tool
in unraveling the mystery of what is happening at super-high scales.

\section*{Appendix}

In this appendix we consider values of $r$ that can arise in fully 
unified models based on unitary groups. 
Consider a model based on $SU(7)$, with each family
consisting of the multiplets $2 \times \overline{[3]} + [2] + [1] =
2 \times \overline{{\bf 35}} + {\bf 21} + {\bf 7}$. The group $SU(7)$
contains the subgroup $SU(5) \times
U(1)_1 \times U(1)_2$, where the generators of the two $U(1)$ groups
are $X_1 = diag (1,1,1,1,1,-\frac{5}{2}, - \frac{5}{2})$ and
$X_2 = diag (0,0,0,0,0,\frac{1}{2}, - \frac{1}{2})$. Each family thus
decomposes into $2 \times (\overline{{\bf 10}}^{3,0} + 
{\bf 10}^{-1/2, 1/2} + {\bf 10}^{-1/2, -1/2} + {\bf 5}^{-4,0})$
$+ (\overline{{\bf 10}}^{-2,0} + \overline{{\bf 5}}^{3/2, -1/2} +
\overline{{\bf 5}}^{3/2, 1/2} + {\bf 1}^{5,0})$ 
$+ (\overline{{\bf 5}}^{-1,0} + {\bf 1}^{5/2, -1/2} + {\bf 1}^{5/2, 1/2})$.
We assume that $SU(5)$ is broken to the Standard Model group at a very high
scale, but we use $SU(5)$ notation to describe the quark and lepton
content for simplicity.
We will consider two breaking schemes in which singlet VEVs first break
the group down to $G_{SM} \times U(1)'$ at a scale $M_*$, and then at
a scale $M' \ll M_*$ other singlets break the $U(1)'$. 

{\bf Case A:} Let Higgs fields in the representations ${\bf 1}^{5/2,1/2}
+{\bf1}^{-5/2,-1/2}$ obtain VEVs of order $M_*$. This breaks $U(1)_1
\times U(1)_2$ down to $U(1)'$, where $X' = X_1 - 5X_2$. The singlet VEVs
also give mass to the pairs of fermions $({\bf 10}^{-1/2,-1/2} +
\overline{{\bf 10}}^{-2,0})$, $2 \times ({\bf 10}^{-1/2,1/2} +
\overline{{\bf 10}}^{3,0})$, and $(\overline{{\bf 5}}^{3/2, -1/2} + 
{\bf 5}^{-4,0})$. That leaves light the following multiplets in each family:
${\bf 10}^{-1/2,-1/2} + \overline{{\bf 5}}^{-1.0} + 
\overline{{\bf 5}}^{3/2,1/2} + {\bf 5}^{-4,0}$. Or in terms of the
$U(1)'$ charges, these are 
${\bf 10}^2 + \overline{{\bf 5}}^{-1} + 
\overline{{\bf 5}}^{-1} + {\bf 5}^{-4}$. This is, in fact, just the same
set of multiplets that arise in $SU(6)$ models, as can be seen by comparing
with the discussion of $SU(6)$ at the beginning of section 2. One sees that
$r=-2$. 

{\bf Case B:} Let Higgs fields in the representations ${\bf 1}^{5,0}
+{\bf1}^{-5,0}$ obtain VEVs of order $M_*$. This breaks $U(1)_1
\times U(1)_2$ down to $U(1)' = U(1)_2$. The singlet VEVs
also give mass to the pair of fermions $(\overline{{\bf 5}}^{-1,0} + 
{\bf 5}^{-4,0})$. At a lower scale, $M'$ the singlets
${\bf 1}^{5/2,1/2} +{\bf1}^{-5/2,-1/2}$ obtain VEVs and give mass
to the pairs $({\bf 10}^{-1/2,-1/2} +
\overline{{\bf 10}}^{-2,0})$, $2 \times ({\bf 10}^{-1/2,1/2} +
\overline{{\bf 10}}^{3,0})$, and $(\overline{{\bf 5}}^{3/2, -1/2} + 
{\bf 5}^{-4,0})$. That leaves light the following multiplets in each family:
${\bf 10}^{-1/2,-1/2} + \overline{{\bf 5}}^{3/2,1/2}$. Or in terms of the
$U(1)'$ charges, these are 
${\bf 10}^{-1/2} + \overline{{\bf 5}}^{1/2}$, implying that $r = -1$.
However, it is clear that Case B is completely unrealistic as a model with a 
low-energy $Z'$ boson, since so many multiplets of quarks and leptons have mass
of order $M'$ that unless $M'$ is near the unification scale the gauge
couplings will blow up below the unification scale. 

This seems to be a general feature in fully unified models based
on unitary groups: either one breaks to an $SU(6)$-like 
low-energy model, giving $r=-2$, or there end up being too many light
quarks and leptons for unification of gauge coupling, i.e. there is a Landau
pole below the unification scale.

\section*{Acknowledgements}

We thank T. Rizzo for calling our attention to the small 
radiatively-induced mixing with hypercharge that had been neglected
in our original analysis. This work was partially supported by the DOE
under the contract No. DE-FG02-91ER40626.

\end{document}